\documentclass[12pt]{article}

\usepackage[margin=1in]{geometry}
\usepackage{setspace}
\usepackage{cite}
\usepackage{amsmath,amssymb,amsfonts}
\usepackage{algorithmic}
\usepackage{graphicx}
\usepackage{textcomp}
\usepackage{siunitx}
\usepackage{lineno}
\usepackage{titlesec}
\usepackage{authblk}
\usepackage{float}
\usepackage{placeins}
\usepackage{multirow}
\usepackage{multicol}
\usepackage{comment}
\titleformat{\section}{\large\bfseries}{\thesection}{0.6em}{}
\titleformat{\subsection}{\normalsize\bfseries}{\thesubsection}{0.6em}{}

\title{Thermal Processing Limits in Oxide-Channel Ferroelectric Field Effect Transistors}
\author{Lance Fernandes*, Yu-Hsin Kuo*,  Chengyang Zhang, Priyankka Ravikumar, Ranie Jeyakumar, Dyutimoy Chakraborty, Jiayi Chen, Taeyoung Song, Kai Ni,  Woohyun Hwang, Kwangyou Seo, Suhwan Lim, Wanki Kim, Daewon Ha, Julia Medvedeva, Suman Datta, Shimeng Yu,  and Asif Khan
\thanks{*L.F and Y.H.K contributed equally to this work}
\thanks{This work was supported by Samsung Electronics (IO250304-12193-01), DOE, EERE, SETO (grant DE-EE0009346), NSF-MRI (grant OAC-1919789). Fab was done at IMS under NSF NNCI (ECCS-1542174). }
\thanks{L.F, Y.H.K, C.Z, P.R, R.J, D.C, J.C, T.S, S.D, S.Y and A.K are with School of ECE, Georgia Institute of Technology, Atlanta, GA, 30318, USA  (e-mail: (lfernandes33, ykuo65, akhan40)@gatech.edu. }
\thanks{K.N is with University of Notre Dame}
\thanks{J.M is with Missouri University of Science and Technology, Missouri, USA}
\thanks{W.H, K.S, S.L, W.K and D.H are with Semiconductor Research and Development, Samsung Electronics Co., Ltd., South Korea.}
\thanks{S.D and A.K are with School of Material Science and Engineering, Georgia Institute of Technology, Atlanta, GA, 30318, USA}
}

\begin{document}

\maketitle
\begin{abstract}
In this work, we report a systematic study of the impact of  high-temperature post-capping thermal annealing on the memory characteristics of Oxide-semiconductor channel ferroelectric field-effect transistors (OS-FeFETs). Using an identical engineered ferroelectric gate stack—8nm Hf$_{0.5}$Zr$_{0.5}$O$_2$ (HZO) / 3 nm Al$_2$O$_3$ / 8 nm HZO (8/3/8)—and a hybrid capping layer (3 nm HfO$_2$ + 3 nm Al$_2$O$_3$), 10\% Ga-doped In$_2$O$_3$ (IGO)–channel and 4\% W-doped In$_2$O$_3$ (IWO)–channel FeFETs remain functional after annealing at temperatures up to 650 $^\circ$C for durations of up to 30 min and 10 min, respectively; further annealing results in irreversible loss of conduction and device failure. Detailed electrical analysis reveals that the MW enhancement originates from a preferential positive shift in the erased-state threshold voltage, while the programmed-state threshold voltage remains comparatively stable. Grazing-incidence X-ray diffraction measurements further indicate structural evolution in the IWO and IGO oxide channels with increasing annealing temperature, supporting the observed electrical trends.


\end{abstract}

\vspace{0.5em}
\vspace{0.5em}

\section{Introduction}
\label{sec:introduction}
The vertically stacked architecture of three-dimensional (3D) charge-trap flash (CTF) NAND enables high storage density, addressing growing data demands \cite{molas2021advances, yoon2022fundamentals}. However, further vertical scaling introduces challenges such as increased power consumption, slower operation, cell-to-cell interference, and threshold voltage instability \cite{DhaVLSI2022}. In addition, conventional 3D-NAND relies on multiple high-temperature steps—including CTF densification and poly-Si crystallization (~650–800 $^\circ$C)—which constrain vertical integration and reliability \cite{HAOEDL24}. To address these issues, fluorite-structure HfO\textsubscript{2}-based ferroelectric memory with a dielectric insert has emerged as a promising alternative. Ferroelectric 3D NAND (3D-FeNAND) enables low-voltage operation, fast switching, and process compatibility \cite{samsungiedm23, florent2018vertical, kuk2024superior, lanceedl24, venkatesan2025pushing}. Moreover, oxide-semiconductor (OS) channels offer strong architectural compatibility for next-generation 3D-NAND \cite{yoo2025ferroelectric, joh2024oxide, fernandes2025comparative}. Replacing CTF and poly-Si with ferroelectric gate stacks and OS channels reduces thermal budget and simplifies processing. However, OS channels remain highly sensitive to thermal processing. In practice, post-channel v-NAND steps impose cumulative anneals exceeding 400 $^\circ$C and approaching ~550 $^\circ$C, yet the thermal limits of oxide-channel FeFETs under such conditions remain largely unexplored.

Here, we report that increasing annealing temperature and duration enhances the memory window in oxide-channel FeFETs via a preferential shift in the erased-state threshold voltage, reaching a maximum just before channel failure (loss of channel conduction). Using a laminated 8 nm HZO / 3 nm Al$_2$O$_3$ / 8 nm HZO stack, IGO and IWO devices remain functional up to 650 $^\circ$C for 30 min and 10 min, respectively. GI-XRD reveals distinct thermally induced structural evolution in the channels, explaining the faster failure of IWO compared to IGO. This study focuses on oxide-channel devices with post-channel capping for high-temperature processing.

\section{Experimental Details}

\begin{figure}[htpb]
\centerline{\includegraphics[width=\columnwidth]{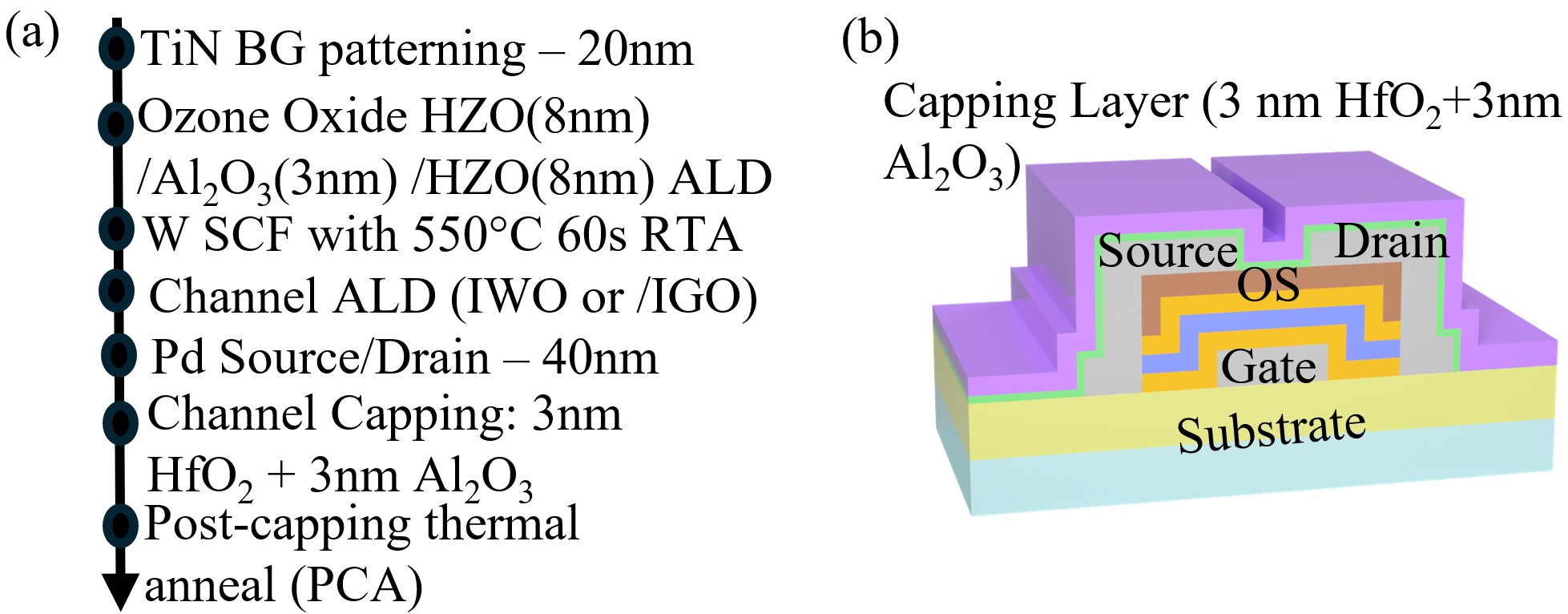}}
\caption{ (a) Fabrication process flow for back-gated OS-FeFET with laminated stack and post-channel capping. (b) 3D schematic of OS-FeFET with capping. }
\label{Fig1}
\end{figure}

Fig. \ref{Fig1}a shows the fabrication flow of back-gated OS-FeFETs with a laminated gate stack deposited by ozone ALD at 250 $^\circ$C. The OS channel (~3.7 nm) is deposited via ALD, with Ga and W concentrations of 10\% and 4\% in In$_2$O$_3$, respectively. After contact formation, a 3 nm HfO$_2$ + 3 nm Al$_2$O$_3$ capping layer is deposited by ozone ALD \cite{kuoted26}. Fig. \ref{Fig1}b shows the corresponding 3D schematic. The capped devices are then annealed under various conditions to evaluate thermal survivability under 3D-NAND–relevant thermal budgets. PCA was performed in N$_2$ at 400–650 $^\circ$C for 10–40 min (50 $^\circ$C/s ramp), followed by natural cooling in N$_2$.

\section{Results and Discussion}


\begin{figure}[htpb]
\centerline{\includegraphics[width=\columnwidth]{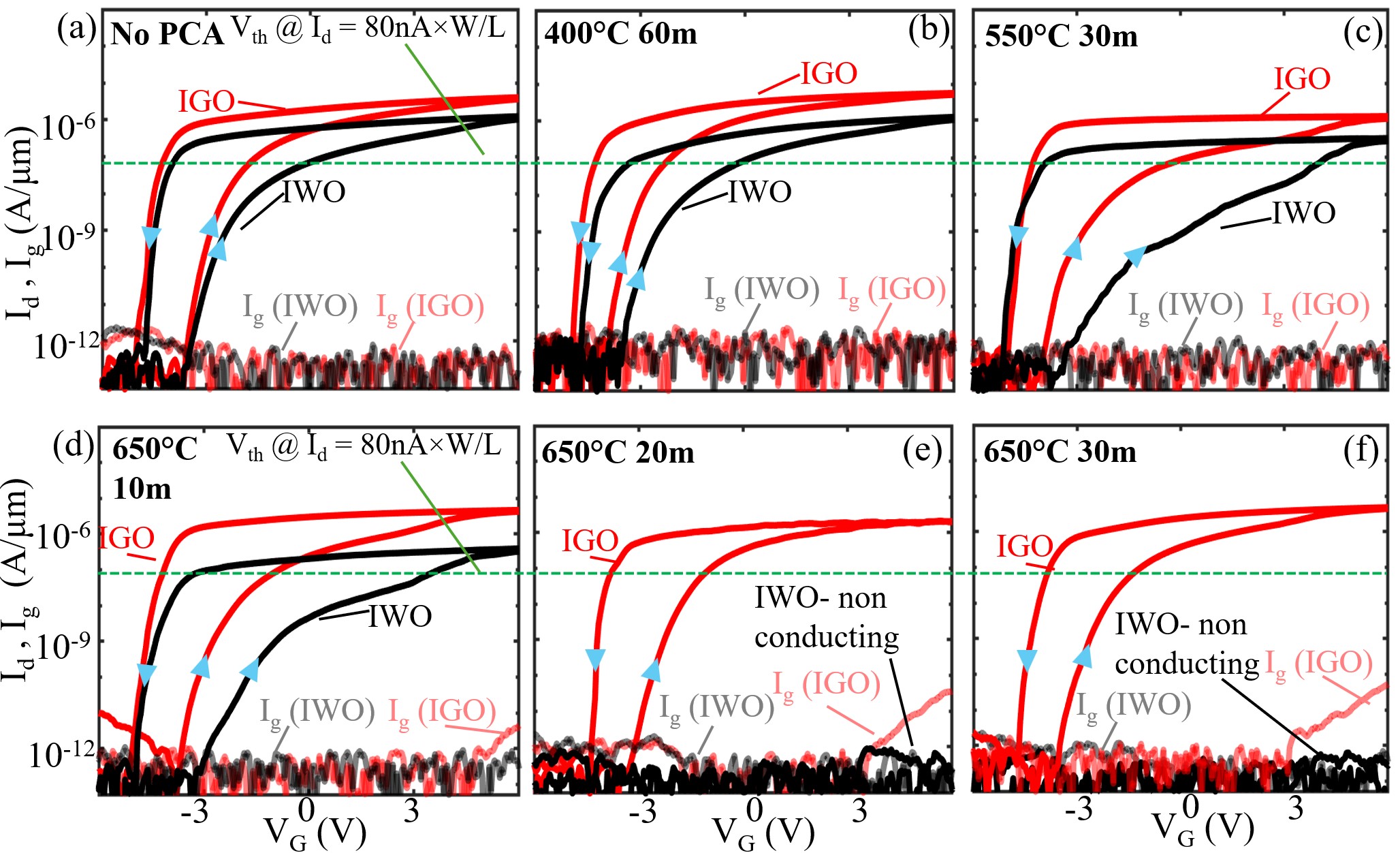}}
\caption{ (a) DC I$_d$-V$_g$ and I$_g$-V$_g$ characteristics of the device without PCA. (b-c) DC I$_d$-V$_g$ and I$_g$-V$_g$ characteristics following moderate-temperature annealing (400–550 $^\circ$C) (d-f) DC I$_d$-V$_g$ and I$_g$-V$_g$ characteristics after high-temperature annealing (650 $^\circ$C). Blue arrows indicate the direction of I-V sweep.}
\label{Fig2}
\end{figure}

Fig. \ref{Fig2}a shows the DC I$_d$-V$_g$  (with anticlockwise ferroelectric hysteresis) and I$_g$-V$_g$ of the control (no PCA), measured at 50 mV drain bias with V$_{th}$ extracted at 80 nA $\times$ W/L. Both IGO and IWO devices remain functional after 400 $^\circ$C PCA (60 min) with negligible ON-current change (Fig. \ref{Fig2}b). At 550 $^\circ$C (30 min), the ON current reduces and the erased-state V$_{th}$ shifts positively, increasing the memory window (MW), while the programmed state remains unchanged (Fig. \ref{Fig2}c); longer anneals lead to channel failure. Gate leakage remains largely unchanged up to 550 $^\circ$C. Fig. \ref{Fig2}(d–f) shows maximum thermal budgets of 650 $^\circ$C/30 min (IGO) and 650 $^\circ$C/10 min (IWO), beyond which conduction is lost, with gate leakage increasing after 20 min at 650 $^\circ$C. The maximum thermal budget is defined as the highest annealing condition retaining channel conduction, extractable V$_{th}$, and a distinguishable memory window. Overall, higher PCA reduces ON current and shifts erased-state V$_{th}$ positively, enhancing MW.

\begin{figure}[htpb]
\centerline{\includegraphics[scale=0.34]{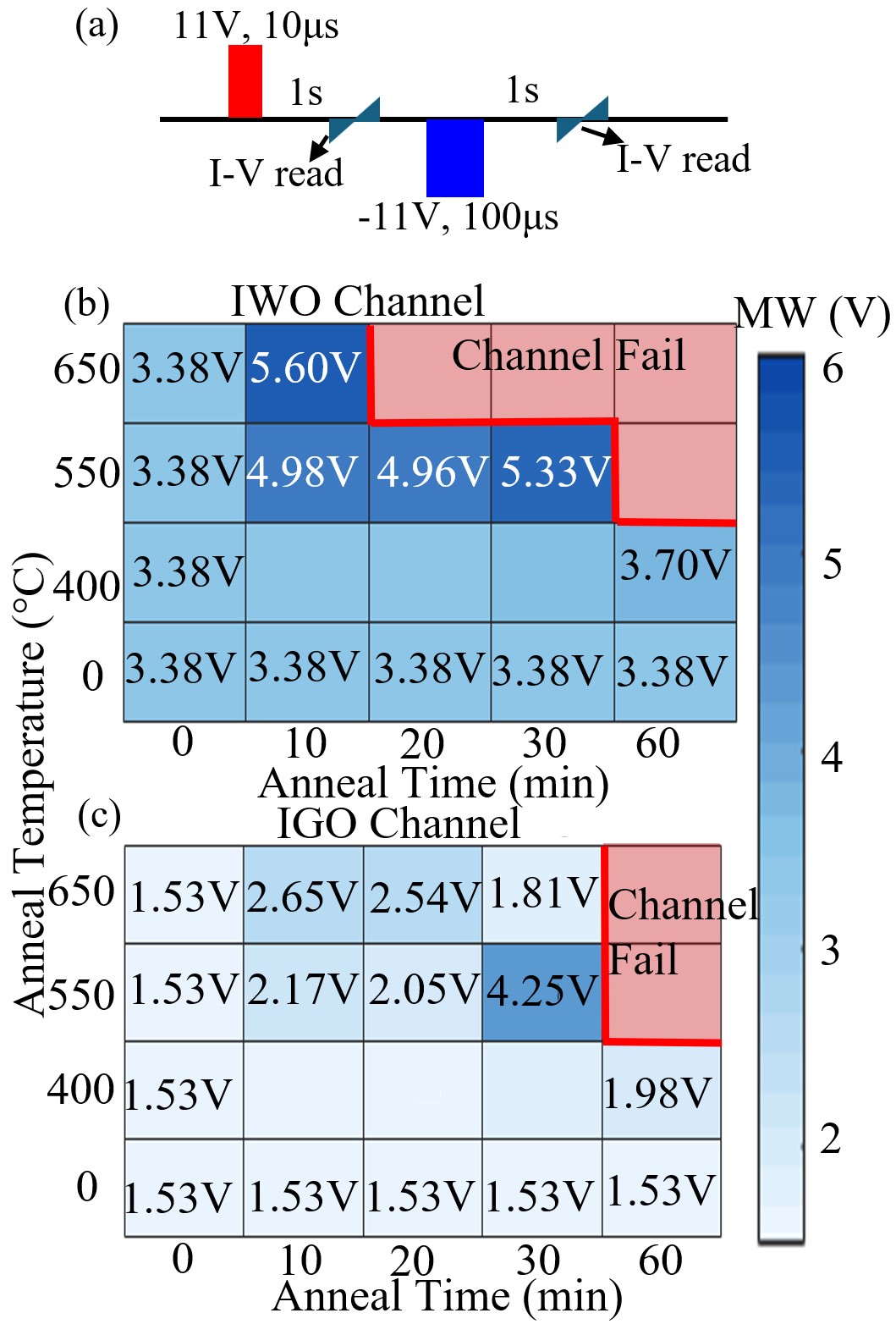}}
\caption{(a) Pulsed I-V measurement scheme used to extract MW. (b-c) Memory window map for IWO and IGO FeFETs for different post-capping annealing conditions (Unlabeled boxes are interpolated from measured data for visualization purposes). Data is averaged over 4-5 devices with a standard deviation of 0.16-0.39}
\label{Fig3}
\end{figure}
Fig. \ref{Fig3}(b-c) presents thermal MW maps for IWO- and IGO-channel FeFETs, illustrating the combined impact of annealing 
temperature and duration. They reveal that MW increases with thermal budget and reaches a pronounced maximum immediately before the onset of thermal-budget-induced channel failure. The MW is extracted using the pulsed I-V scheme shown in Fig. \ref{Fig3}a. Moreover, IWO channel exhibits higher MW than IGO channel for all annealing conditions.

\begin{figure}[htpb]
\centerline{\includegraphics[scale=0.26]{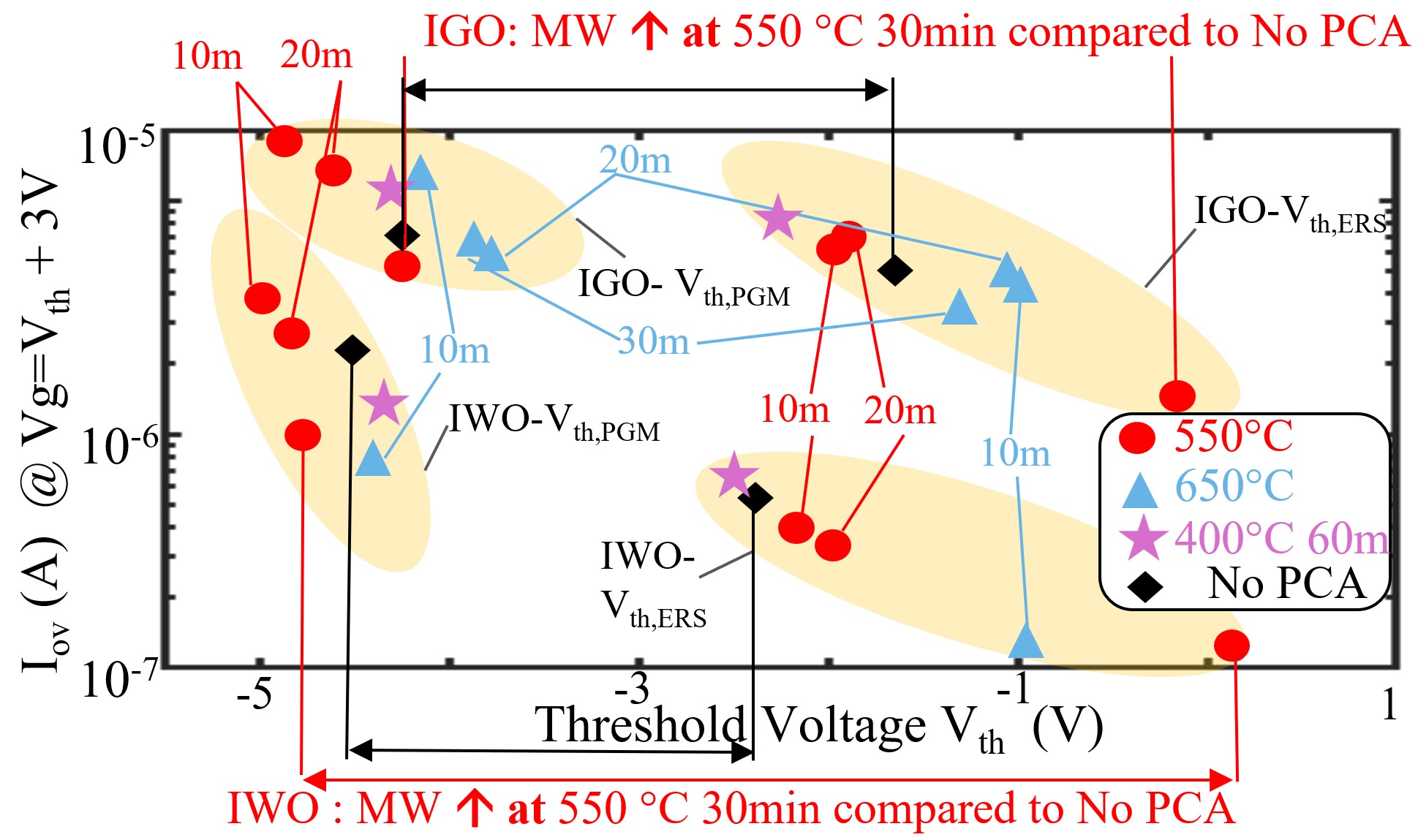}}
\caption{Overdrive current at V$_g$=V$_{th}$+3V (=I$_{ov}$) as a function of V$_{th}$ for PGM and ERS states (V$_{t,PGM}$ and V$_{t,ERS}$) for all annealing conditions.}
\label{Fig4}
\end{figure}

To understand the origin of MW enhancement, Fig. \ref{Fig4} plots I$_{ov}$ as a function of programmed and erased threshold voltages under different annealing conditions. With increasing thermal budget, V$_{th,ERS}$ shows a larger positive shift than V$_{th,PGM}$, resulting in increased MW (=V$_{th,ERS}$- V$_{th,PGM}$). Concurrently, I$_{ov}$ decreases monotonically with increasing V$_{th}$ for both states, consistent with the inverse dependence of carrier concentration on threshold voltage in oxide-semiconductor TFTs \cite{pgr_wang2015analysis}. In conventional OS-MOSFETs, such positive V$_{th}$ shifts are typically associated with increased effective series resistance \cite{pgr_shimura2008specific,pgr_wang2015analysis}. Increasing the thermal budget, lowers the Oxygen vacancies in channel, which leads to higher channel and contact resistance in the oxide-semiconductor channel. In FeFETs, this effect primarily impacts the high-V$_{th}$ erased state, where the depleted channel places the device in an injection-limited regime, making V$_{th}$ sensitive to channel and contact resistance. Higher contact resistance due to depleted channel makes electron injection more difficult, shifting the high-V$_{th}$ state \cite{10413793}. Similarly, the subthreshold swing (SS) degrades with increasing PCA as seen in Fig. \ref{Fig2} due to an increase in contact resistance \cite{10413793}, which is also observed in standard OS-MOSFETs \cite{lee2010systematic}.  The SS degradation is more pronounced in the erased state, consistent with an injection-limited transport regime. The SS degradation under the maximum thermal budget (prior to channel failure) in the erased state is 1.74$\times$ for the IWO channel (650 $^\circ$C/10 min) and 1.41$\times$ for the IGO channel (650 $^\circ$C/30 min) compared to the no-PCA case. In contrast, the low-V$_{th}$ state is dominated by ferroelectric polarization, which induces strong channel accumulation and reduces the effective energy barrier at the contacts \cite{10413793}. Consequently, even with increased contact resistance after PCA, carrier injection remains efficient, preventing it from becoming the dominant limiting factor as in the high-V$_{th}$ state.

\begin{figure}[htpb]
\centerline{\includegraphics[width=\columnwidth]{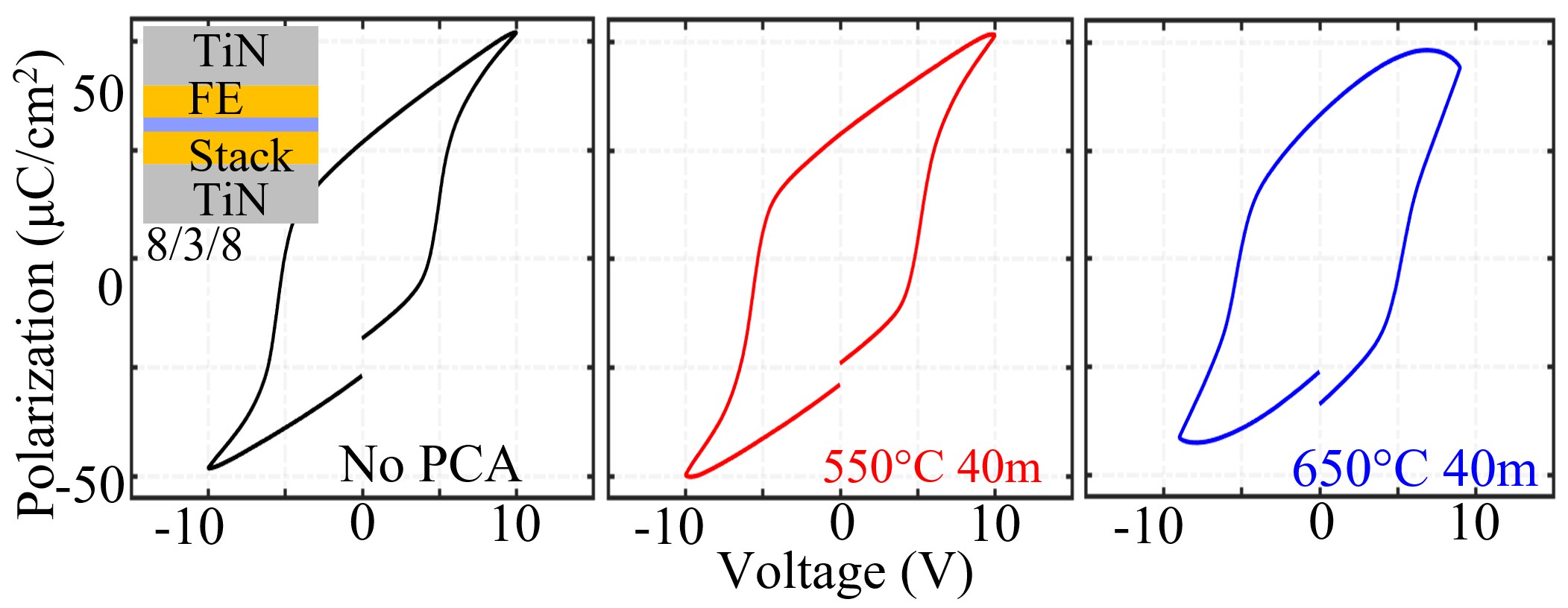}}
\caption{Polarization vs. Voltage characteristics on MFM capacitor for before and after PCA at 550 $^\circ$C and 650 $^\circ$C beyond the channel failure time.}
\label{Fig5a}
\end{figure}

Moreover, P–V characterization of the laminated MFM stack (Fig. \ref{Fig5a}) shows robust ferroelectric switching with minimal change in coercive voltage and remanent polarization even beyond channel failure at 550 $^\circ$C and 650 $^\circ$C PCA. We note that increased leakage is observed at 650 $^\circ$C for 40 min. This indicates that MW evolution is primarily governed by oxide-channel transport rather than changes in ferroelectric properties.

\begin{figure}[htpb]
\centerline{\includegraphics[width=\columnwidth]{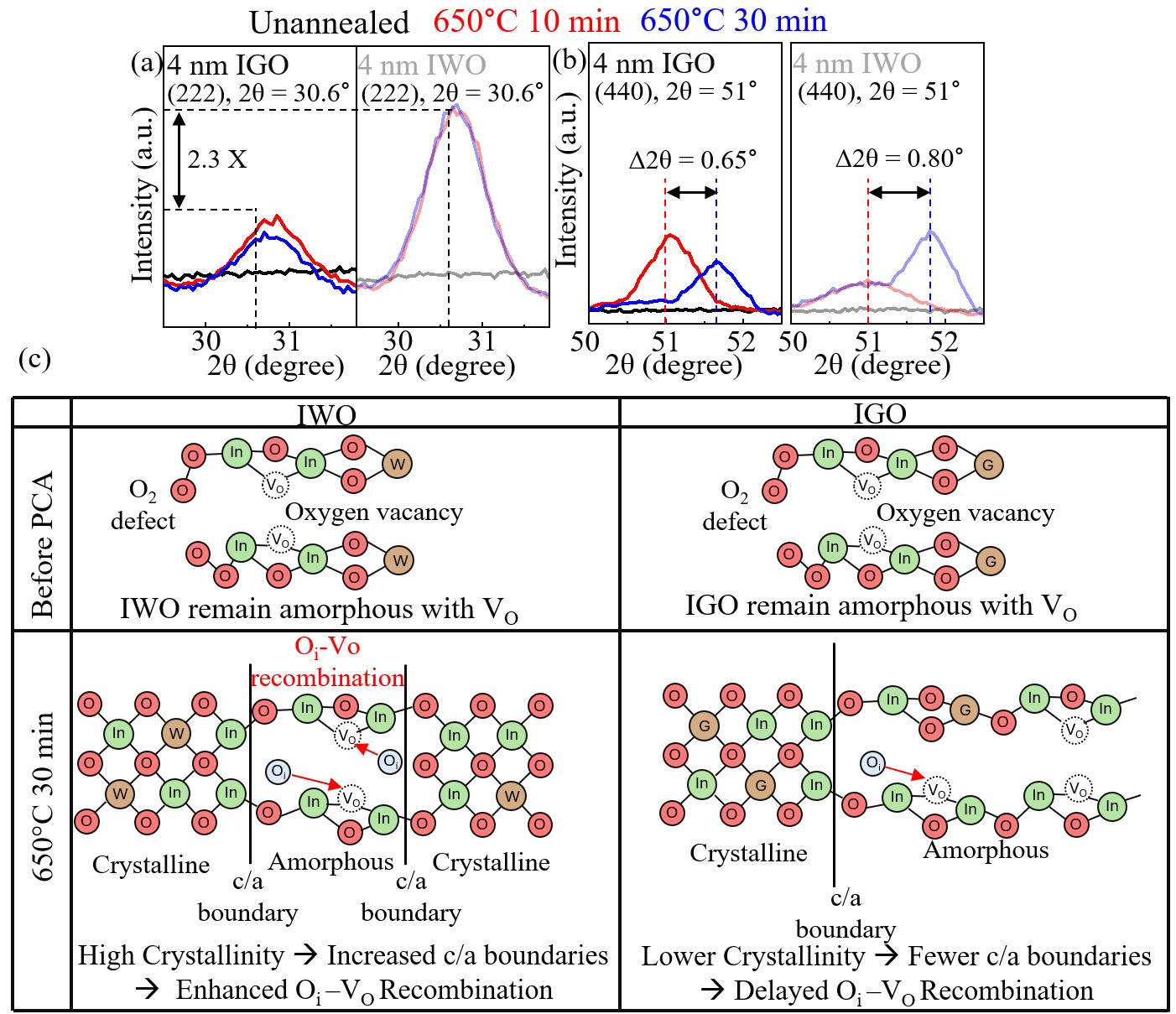}}
\caption{Zoomed-in GI-XRD post-capping for unannealed and 650 $^\circ$C PCA conditions for (a) (222) and (b) (440) peaks (c) Pictorial representation of the differences between IWO and IGO channels.}
\label{Fig5}
\end{figure}

At high processing temperatures and durations, O$_2$ bonds dissociate, generating O$_i$ that recombines with Oxygen vacancies (V$_O$), quenching vacancy donors and eliminating channel conduction, consistent with experimental observations. The longer time-to-failure in IGO (40 min) compared to IWO (20 min) is attributed to morphology-driven defect proximity. GI-XRD reveals a 2.3$\times$ stronger (222) peak in IWO (Fig. \ref{Fig5}a), indicating higher crystallinity and reduced amorphous regions (Fig. \ref{Fig5}c). Since V$_O$ preferentially resides (and is more mobile) in the amorphous phase \cite{medvedeva2025elucidating_1}, while O$_2$ linkages are pinned at crystalline/amorphous (c/a) boundaries \cite{medvedeva2022hydrogen_2}, higher crystallinity in IWO reduces O$_2$–V$_O$ separation and accelerates recombination. In contrast, the larger amorphous fraction in IGO increases this separation, delaying conductivity loss (Fig. \ref{Fig5}c). This behavior is driven by high-temperature O$_2$ bond dissociation and subsequent V$_O$ recombination, indicating that channel degradation is thermal rather than crystallization-induced. The (440) peak remains unshifted at 10 min ($\simeq$51$^o$), consistent with near–dopant-free bixbyite, but shifts to higher 2$\theta$ after 30 min (Fig. \ref{Fig5}b), indicating W/Ga incorporation and lattice contraction (ionic radii $\leq$ In). Notably, the (440) intensity is higher in IGO at 10 min but lower at 30 min, suggesting that Ga incorporation both contracts the lattice and introduces crystallographic disorder.

\begin{figure}[htpb]
\centerline{\includegraphics[width=\columnwidth]{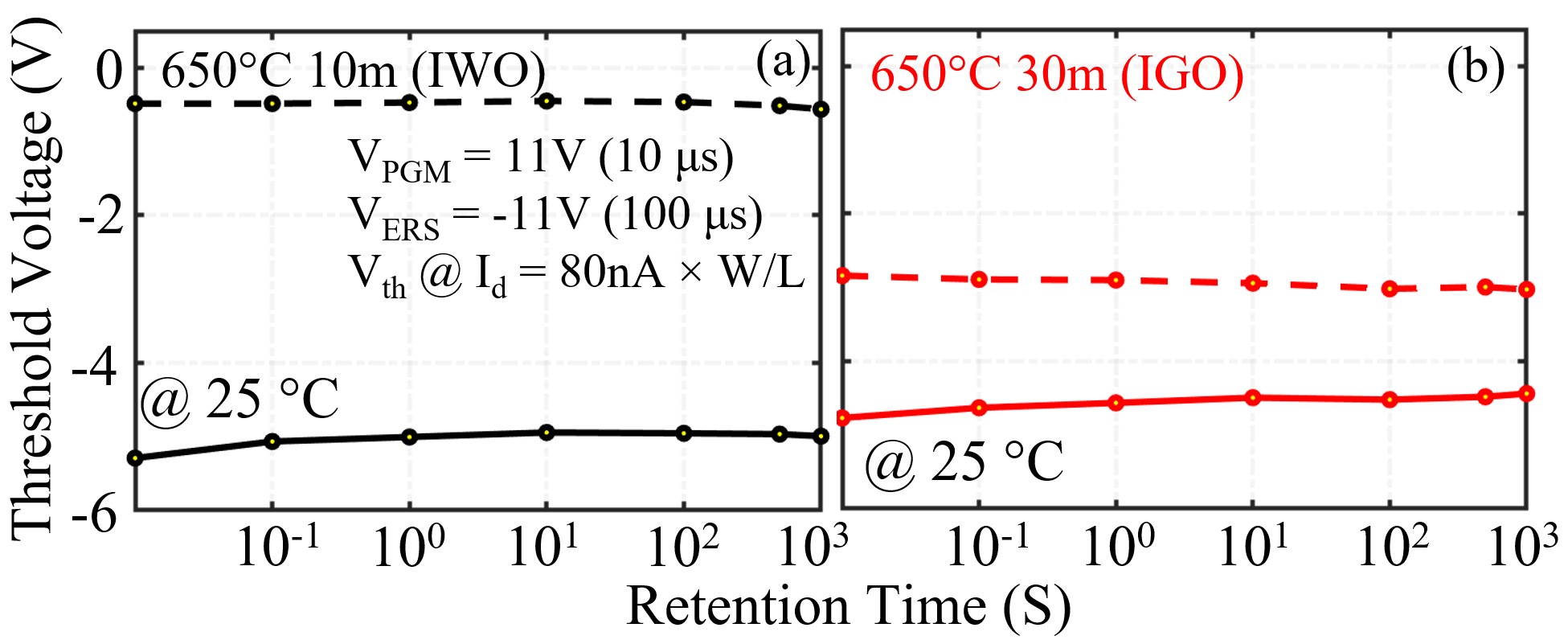}}
\caption{Retention at room temperature under respective maximum thermal budget conditions for (a) IWO and (b) IGO channel.}
\label{Fig7}
\end{figure}

Fig. \ref{Fig7}(a-b) shows retention characteristics of IWO and IGO channel FeFETs after PCA at respective maximum thermal budget showing robust programmed and erased states even after PCA

\section{Conclusion}

This work systematically establishes oxide channel FeFETs as 3D-NAND viable by demonstrating relevant thermal compatibility and material-dependent survivability. Increasing the thermal budget increase the memory window and reduces the ON current in both IWO and IGO channel with IWO channel exhibiting higher MW than IGO channel. Grazing-incidence X-ray diffraction measurements reveal thermally induced structural evolution in the oxide channels, which explains the faster channel degradation observed in IWO compared to IGO. These results provide important insights into the relationship between thermal processing and memory characteristics in oxide-channel FeFETs and offer guidance for optimizing thermal budgets in FeFET-based 3D-NAND integration.

\bibliographystyle{IEEEtran}
\bibliography{ref}

\end{document}